\documentclass{ws-ijmpcs}

\begin{document}

\markboth{A. A. Saharian} {Quantum Effects in Higher-Dimensional
Cosmological Models}

%
\catchline{}{}{}{}{}
%

\title{\bf QUANTUM EFFECTS IN HIGHER-DIMENSIONAL COSMOLOGICAL MODELS
}

\author{ARAM A. SAHARIAN\footnote{saharian@ysu.am}}

\address{Department of Physics, Yerevan State University\\
1 Alex Manoogian Street, 0025 Yerevan, Armenia}

\maketitle

\begin{history}
\received{Day Month Year} \revised{Day Month Year}
\end{history}

\begin{abstract}
Vacuum energy density and stresses are investigated for a scalar
field in de Sitter spacetime with an arbitrary number of
toroidally compactified spatial dimensions and in anti-de Sitter
spacetime with two parallel branes. On the branes the field obeys
the Robin boundary conditions. The behavior of the vacuum
expectation values is discussed in various asymptotic regions of
the parameters. Applications are given to Randall-Sundrum type
braneworlds.

\end{abstract}

\ccode{PACS numbers: 04.62.+v, 04.50.-h, 11.10.Kk, 04.20.Gz}

\section{Introduction}

A number of high-energy theories of fundamental physics are
formulated in higher-dimensional spacetimes. In particular, the
idea of extra dimensions has been used in supergravity and
superstring theories. Two types of models with extra dimensions
were discussed in the literature. The first one corresponds to
Kaluza-Klein type models, where it is assumed that the extra
dimensions are compactified and all of them are accessable for the
standard model fields. In the second type of models, the standard
model fields are localized on a 3-dimensional hypersurface (brane)
and the extra dimensions are accessable for the gravitational and
for some additional hypothetical fields. These models provide a
solution to the hierarchy problem between the gravitational and
electroweak mass scales \cite{Arka98,Rand99a}. The main idea to
resolve the large hierarchy is that the small coupling of four
dimensional gravity is generated by the large physical volume of
extra dimensions. These theories provide a novel setting for
discussing phenomenological and cosmological issues related to
extra dimensions.

In both types of models with extra dimensions the boundary
conditions imposed on possible field configurations change the
spectrum of the vacuum fluctuations and lead to the Casimir-type
contributions to the vacuum expectation values of physical
observables. The Casimir effect (for a review see Ref.
\refcite{Milt02}) has been used as a stabilization mechanism for
moduli fields parametrizing the size of the extra dimensions. The
Casimir energy coming from the extra dimensions can also serve as
a model for dark energy (see Ref.~\refcite{Milt03} and references
therein). In the first part of the present talk, based on
Refs.~\refcite{Saha07,Bell08}, we describe the effects of the
toroidal compactification of spatial dimensions in de Sitter (dS)
spacetime on the properties of quantum vacuum for a scalar field
with general curvature coupling parameter. The corresponding
one-loop quantum effects for a fermionic field are studied in
Refs.~\refcite{Saha08,Beze08}. The second part, based on Refs.
\refcite{Saha05,Saha06}, is devoted to the investigation of the
vacuum energy-momentum tensor for a scalar field in the geometry
of two parallel branes in anti-de Sitter (AdS) spacetime. This
geometry is employed in the Randall-Sundrum type braneworld models \cite%
{Rand99a}.

\section{dS spacetime with totoidally compact dimensions}

\label{sec:Geom}

Consider a free scalar field with curvature coupling parameter
$\xi $. The corresponding field equation has the form
\begin{equation}
\left( \nabla _{l}\nabla ^{l}+m^{2}+\xi R\right) \varphi =0,
\label{fieldeq}
\end{equation}%
where $R$ is the Ricci scalar for the background spacetime. For
minimally and conformally coupled scalars one has $\xi =0$ and
$\xi =\xi _{c}=(D-1)/4D$ correspondingly. In this section, as a
background geometry we consider $(D+1)
$-dimensional dS spacetime generated by a positive cosmological constant $%
\Lambda $. In planar coordinates the corresponding line element
has the form
\begin{equation}
ds^{2}=dt^{2}-e^{2t/\alpha }\sum_{l=1}^{D}(dx^{l})^{2},
\label{ds2deSit}
\end{equation}%
with the parameter $\alpha ^{2}=D(D-1)/(2\Lambda )$. For this geometry $%
R=D(D+1)/\alpha ^{2}$. We assume that the spatial coordinate $x^{l}$, $%
l=p+1,\ldots ,D$, is compactified to $\mathrm{S}^{1}$ of the
length $L_{l}$:
$0\leqslant x^{l}\leqslant L_{l}$, and for the other coordinates we have $%
-\infty <x^{l}<+\infty $, $l=1,\ldots ,p$. Hence, we consider the
spatial topology $\mathrm{R}^{p}\times (\mathrm{S}^{1})^{q}$,
where $q=D-p$. Along the compact dimensions we impose the
periodicity conditions
\begin{equation}
\varphi (t,x^{1},\ldots ,x^{p},x^{p+1},\ldots ,x^l+L_l,\ldots
,x^{D})=\pm \varphi (t,x^{1},\ldots ,x^{p},x^{p+1},\ldots
,x^l,\ldots ,x^{D}),  \label{periodicBC}
\end{equation}%
where $l=p+1,\ldots ,D$, and the upper/lower sign corresponds to
untwisted/twisted scalar field.

As a result of boundary conditions, the spectrum of the vacuum
fluctuation of the field $\varphi $ is modified compared with the
case of dS spacetime with trivial topology $\mathrm{R}^{D}$. This
leads to the change of the vacuum expectation values (VEVs) for
physical observables (topological Casimir effect). We are
interested in the VEV of the energy-momentum tensor. This quantity
acts as a source of gravity in the semiclassical Einstein
equations and plays an important role in modelling self-consistent
dynamics involving the gravitational field. Assuming that the
field is prepared in the Bunch-Davies vacuum state \cite{Birr82},
we have the following
recurrence relation for the VEV of the energy-momentum tensor%
\begin{equation}
\langle T_{i}^{k}\rangle _{p,q}=\langle T_{i}^{k}\rangle
_{p+1,q-1}+\Delta _{p+1}\langle T_{i}^{k}\rangle _{p,q}.
\label{TikDecomp}
\end{equation}%
Here $\langle T_{i}^{k}\rangle _{p+1,q-1}$ is the VEV for dS
spacetime with
topology $\mathrm{R}^{p+1}\times (\mathrm{S}^{1})^{q-1}$ and the part $%
\Delta _{p+1}\langle T_{i}^{k}\rangle _{p,q}$ is induced by the
compactness
along the $x^{p+1}$ - direction.

For the corresponding energy density one has%
\begin{eqnarray}
\Delta _{p+1}\langle T_{0}^{0}\rangle _{p,q} &=&\frac{2\eta ^{D}L_{p+1}}{%
(2\pi )^{(p+3)/2}V_{q}\alpha ^{D+1}}\sum_{n=1}^{\infty }(\pm 1)^{n}\sum_{%
\mathbf{n}_{q-1}}\int_{0}^{\infty }dx  \notag \\
&&\times \frac{xF_{\nu }^{(0)}(x\eta )}{(nL_{p+1})^{p-1}}f_{(p-1)/2}(nL_{p+1}%
\sqrt{x^{2}+k_{\mathbf{n}_{q-1}}^{2}}),  \label{DelT00}
\end{eqnarray}%
where $V_{q}=L_{p+1}\cdots L_{D}$ is the volume of the compact subspace, $%
\sum_{\mathbf{n}_{q-1}}=\sum_{n_{p+2}=-\infty }^{+\infty }\cdots
\sum_{n_{D}=-\infty }^{+\infty }$,  $f_{\mu }(y)=y^{\mu }K_{\mu }(y)$, $%
K_{\mu }(y)$ is the MacDonald function and%
\begin{eqnarray}
F_{\nu }^{(0)}(y) &=&y^{2}\left[ I_{-\nu }^{\prime }(y)+I_{\nu }^{\prime }(y)%
\right] K_{\nu }^{\prime }(y)+D(1/2-2\xi )yG_{\nu }^{\prime }(y)  \notag \\
&&+G_{\nu }(y)\left( \nu ^{2}+2m^{2}\alpha ^{2}-y^{2}\right) ,
\label{F0}
\end{eqnarray}%
with $I_{\nu }(y)$ being the modified Bessel function, $\nu =%
\sqrt{D^{2}/4-D(D+1)\xi -m^{2}\alpha ^{2}}$ and%
\begin{equation}
G_{\nu }(y)=\left[ I_{-\nu }(y)+I_{\nu }(y)\right] K_{\nu }(y).
\label{nu}
\end{equation}%
In (\ref{DelT00}), the upper/lower sign corresponds to
untwisted/twisted scalar field,
\begin{equation}
k_{\mathbf{n}_{q-1}}^{2}=\sum_{l=p+2}^{D}[2\pi
(n_{l}+g_{l})/L_{l}]^{2}, \label{knq}
\end{equation}%
where $g_{l}=0$ for untwisted scalar and $g_{l}=1/2$ for twisted
scalar
field.

The vacuum stresses are presented in the form (no summation over $i$)%
\begin{eqnarray}
\Delta _{p+1}\langle T_{i}^{i}\rangle _{p,q} &=&\frac{2\eta ^{D}L_{p+1}}{%
(2\pi )^{(p+3)/2}V_{q}\alpha ^{D+1}}\sum_{n=1}^{\infty }\frac{(\pm 1)^{n}}{%
(nL_{p+1})^{p-1}}\sum_{\mathbf{n}_{q-1}}\int_{0}^{\infty }dx\,x  \notag \\
&& \times \Big[ \,F_{\nu }(x\eta )f_{(p-1)/2}(nL_{p+1}\sqrt{x^{2}+k_{\mathbf{%
n}_{q-1}}^{2}}) \nonumber \\
&& -\frac{2G_{\nu }(x\eta )}{(nL_{p+1}/\eta )^{2}}%
f_{p}^{(i)}(nL_{p+1}\sqrt{x^{2}+k_{\mathbf{n}_{q-1}}^{2}})\Big] ,
\label{DelTii1}
\end{eqnarray}%
where we have introduced the notations%
\begin{eqnarray}
f_{p}^{(i)}(y) &=&f_{(p+1)/2}(y),\;i=1,\ldots
,p,\;f_{p}^{(p+1)}(y)=-y^{2}f_{(p-1)/2}(y)-pf_{(p+1)/2}(y),  \notag \\
f_{p}^{(i)}(y) &=&[2\pi
n(n_{i}+g_{i})L_{p+1}/L_{i}]^{2}f_{(p-1)/2}(y),\;i=p+2,\ldots ,D,
\label{fp+1} \\
F_{\nu }(y) &=&\left[ 2(D+1)\xi -D/2\right] yG_{\nu }^{\prime
}(y)+\left( 4\xi -1\right) y^{2}K_{\nu }^{\prime }(y)[ I_{-\nu
}^{\prime }(y)+I_{\nu }^{\prime
}(y)]  \notag \\
&& +G_{\nu }(y)\left[ \left( 4\xi -1\right) \left( y^{2}+\nu
^{2}\right) \right] . \label{Fnu}
\end{eqnarray}%
As it is seen from the obtained formulas, the topological parts in
the VEVs are time-dependent and, hence, the local dS symmetry is
broken by the non-trivial topology. By taking into account the
relation between the conformal and synchronous time coordinates,
we see that the VEV of the energy-momentum tensor is a function of
the combination $L_{l}/\eta =L_{l}e^{t/\alpha }/\alpha $, which is
the comoving length of the compactified dimension measured in
units of the dS curvature scale.

The recurring application of formula (\ref{TikDecomp}) allows us
to write
the VEV in the form%
\begin{equation}
\langle T_{i}^{k}\rangle _{p,q}=\langle T_{i}^{k}\rangle _{\mathrm{dS}%
}+\langle T_{i}^{k}\rangle _{\mathrm{c}},\;\langle T_{i}^{k}\rangle _{%
\mathrm{c}}=\sum_{l=1}^{q}\Delta _{D-l+1}\langle T_{i}^{k}\rangle
_{D-l,l}, \label{TikComp}
\end{equation}%
where the part corresponding to uncompactified dS spacetime,
$\langle T_{i}^{k}\rangle _{\mathrm{dS}}$, is explicitly
decomposed. As a consequence of the maximal symmetry for dS
spacetime and the Bunch-Davies vacuum state,
the latter has the structure $\langle T_{ik}\rangle _{\mathrm{dS}}=\mathrm{%
const}\cdot g_{ik}$. The part $\langle T_{i}^{k}\rangle
_{\mathrm{c}}$ is induced by the comactness of the $q$-dimensional
subspace. This part is finite and the renormalization is needed
for the uncompactified dS part only. We could expect this result,
since the local geometry is not changed by the toroidal
compactification.

For a conformally coupled massless scalar field $\nu =1/2$ and we
find (no
summation over $i$)%
\begin{equation}
\Delta _{p+1}\langle T_{i}^{i}\rangle _{p,q}=-\frac{2(\eta /\alpha
)^{D+1}L_{p+1}}{(2\pi )^{p/2+1}V_{q}}\sum_{n=1}^{\infty }(\pm 1)^{n}\sum_{%
\mathbf{n}_{q-1}\in \mathbb{Z}_{q-1}}\frac{g_{p}^{(i)}(nL_{p+1}k_{\mathbf{n}%
_{q-1}})}{(nL_{p+1})^{p+2}},  \label{DelTConf}
\end{equation}%
with the notations%
\begin{eqnarray}
g_{p}^{(i)}(y) &=&f_{p/2+1}(y),\;i=0,1,\ldots
,p,\;g_{p}^{(p+1)}(y)=-(p+1)f_{p/2+1}(y)-y^{2}f_{p/2}(y),  \notag \\
g_{p}^{(i)}(y) &=&[2\pi
n(n_{i}+g_{i})L_{p+1}/L_{i}]^{2}f_{p/2}(y),\;i=p+2,\ldots ,D.
\label{gi}
\end{eqnarray}%
In this case the topological part in the VEV of the
energy-momentum tensor is traceless and the trace anomaly is
contained in the uncompactified dS part only. Formula
(\ref{DelTConf}) could be obtained from the corresponding
result in $(D+1)$-dimensional Minkowski spacetime with spatial topology $%
\mathrm{R}^{p}\times (\mathrm{S}^{1})^{q}$, taking into account
that two problems are conformally related: $\Delta _{p+1}\langle
T_{i}^{k}\rangle
_{p,q}=(\eta /\alpha )^{D+1}\Delta _{p+1}\langle T_{i}^{k}\rangle _{p,q}^{\mathrm{(M)}%
}$.

For small values of the ratio, $L_{p+1}/\eta \ll 1$, to the leading order, $%
\Delta _{p+1}\langle T_{i}^{k}\rangle _{p,q}$ coincides with the
corresponding result for a conformally coupled massless field, given by (\ref%
{DelTConf}). For a fixed value of the ratio $L_{p+1}/\alpha $,
this limit corresponds to $t\rightarrow -\infty $ and the
topological part $\langle T_{i}^{k}\rangle _{\mathrm{c}}$ behaves
like $\exp [-(D+1)t/\alpha ]$. By taking into account that the
part $\langle T_{i}^{k}\rangle _{\mathrm{dS}}$ is time
independent, from here we conclude that in the early stages of the
cosmological expansion the topological part dominates in the VEV\
of the energy-momentum tensor. In particular, in this limit the
total energy density is negative.

For small values of the ratio $\eta /L_{p+1}$ and for real values
of the parameter $\nu $ we have $\Delta _{p+1}\langle
T_{0}^{0}\rangle _{p,q}\propto (\eta /L_{p+1})^{D-2\nu }$. The
corresponding energy density is positive for a minimally coupled
scalar field and for a conformally coupled massive scalar field.
For the vacuum stresses, to the leading order, we have the
relation (no summation over $i$) $\Delta _{p+1}\langle
T_{i}^{i}\rangle _{p,q}\approx (2\nu /D)\Delta _{p+1}\langle
T_{0}^{0}\rangle _{p,q}$, $i=1,\ldots ,D$, for $\eta /L_{p+1}\ll
1$. The coefficient in this relation does not depend on $p$ and it
takes place for the total topological part of the VEV as well.
Hence, in the limit under consideration the topological parts in
the vacuum stresses are isotropic. Note that this limit
corresponds to late times in terms of synchronous time coordinate
$t$, $(\alpha /L_{p+1})e^{-t/\alpha }\ll 1$, and the topological
part in the VEV is suppressed by the factor $\exp [-(D-2\nu
)t/\alpha ]$. As the uncompactified dS part is constant, it
dominates the topological part at the late stages of the
cosmological evolution.

For small values of the ratio $\eta /L_{p+1}$ and for purely imaginary $\nu $%
, the energy density and the vacuum stresses behave like $\Delta
_{p+1}\langle T_{0}^{0}\rangle _{p,q}\propto e^{-Dt/\alpha }\sin
\left( 2|\nu |t/\alpha +\phi \right) $ and $\Delta _{p+1}\langle
T_{i}^{i}\rangle _{p,q}\propto e^{-Dt/\alpha }\cos \left( 2|\nu
|t/\alpha +\phi \right) $. Hence, in the case under consideration,
at late stages of the cosmological evolution the topological part
is suppressed by the factor $\exp (-Dt/\alpha )$ and the damping
of the corresponding VEV is oscillatory.

\section{Vacuum densities for branes in anti-de Sitter spacetime}

\label{sec:WF}

In this section as a background geometry we consider
$(D+1)$-dimensional AdS spacetime. This spacetime is one of the
simplest and most interesting backgrounds allowed by general
relativity. In Poincare coordinates the corresponding line element
is given by
\begin{equation}
ds^{2}=g_{ik}dx^{i}dx^{k}=e^{-2y/\alpha }\eta _{\mu \nu }dx^{\mu
}dx^{\nu }-dy^{2},  \label{metric}
\end{equation}%
where $\eta _{\mu \nu }$ is the metric for the $D$-dimensional
Minkowski spacetime. For the corresponding Ricci scalar one has
$R=-D(D+1)/\alpha ^{2}$. By making a coordinate transformation
$z=\alpha e^{y/\alpha }$, metric (\ref{metric}) is written in a
conformally-flat form: $ds^{2}=(\alpha /z)^{2}\eta
_{ik}dx^{i}dx^{k}$.

Below we will study quantum vacuum effects induced by two parallel
infinite branes, located at $y=a$ and $y=b$, $a<b$. On the branes
the field $\varphi $ obeys the Robin boundary conditions
\begin{equation}
(\tilde{A}_{y}+\tilde{B}_{y}\partial _{y})\varphi (x)=0,\quad
y=a,b, \label{boundcond}
\end{equation}%
with constant coefficients $\tilde{A}_{y}$, $\tilde{B}_{y}$. The
presence of boundaries modifies the spectrum for the zero--point
fluctuations of the scalar field under consideration. This leads
to the modification of the VEVs of physical quantities compared
with the case of AdS spacetime without branes. In particular,
vacuum forces arise acting on the boundaries. Motivated by the
applications to braneworld type scenarios, the vacuum energy in
the two-branes setup has been considered in a large number of
papers (see, for instance, references given in Refs.
\refcite{Eliz09}). Here we present the results for the VEV of the
energy-momentum tensor.

\subsection{Casimir densities for a single brane}

\label{sec:CD1b}

In this section we will consider the VEV of the geometry of a
single brane located at $y=a$. The corresponding vacuum
energy-momentum tensor is diagonal and is presented in the form
\begin{equation}
\langle 0|T_{i}^{k}|0\rangle =\langle T_{i}^{k}\rangle _{\mathrm{AdS}%
}+\langle T_{i}^{k}\rangle ^{(a)},  \label{EMT41pl}
\end{equation}%
where $\langle T_{i}^{k}\rangle _{\mathrm{AdS}}$ is the
renormalized VEV for the energy-momentum tensor in the AdS
background without boundaries, and the
term $\langle T_{i}^{k}\rangle ^{(a)}$ is induced by a single brane at $y=a$%
. The brane-induced part is finite for points away the brane and
the renormalization procedure is needed for the boundary-free part
only. The latter is well-investigated in the literature. Due to
the maximal symmetry
of AdS spacetime one has $\langle T_{i}^{k}\rangle _{\mathrm{AdS}}=\mathrm{%
const}\cdot g_{ik}$. The boundary-induced part is given by the
expressions
\begin{eqnarray}
\langle T_{i}^{k}\rangle ^{(a)} &=&-\frac{\alpha ^{-D-1}z^{D}\delta _{i}^{k}%
}{2^{D-1}\pi ^{D/2}\Gamma (D/2)}\int_{0}^{\infty }duu^{D-1}\frac{\bar{I}%
_{\sigma }^{(a)}(uz_{a})}{\bar{K}_{\sigma
}^{(a)}(uz_{a})}F^{(i)}\left[
K_{\sigma }(uz)\right] ,\;z>z_{a},  \notag \\
\langle T_{i}^{k}\rangle ^{(a)} &=&-\frac{\alpha ^{-D-1}z^{D}\delta _{i}^{k}%
}{2^{D-1}\pi ^{D/2}\Gamma (D/2)}\int_{0}^{\infty }duu^{D-1}\frac{\bar{K}%
_{\sigma }^{(a)}(uz_{a})}{\bar{I}_{\sigma
}^{(a)}(uz_{a})}F^{(i)}\left[ I_{\sigma }(uz)\right] ,\;z<z_{a},
\label{Tik1plnew2}
\end{eqnarray}%
where $z_{j}=\alpha e^{j/\alpha }$ and $\sigma
=\sqrt{(D/2)^{2}-D(D+1)\xi +m^{2}\alpha ^{2}}$. In
(\ref{Tik1plnew2}), for a function $F(x)$ we have defined the
barred notation
\begin{equation}
\bar{F}^{(j)}(x)=A_{j}F(x)+B_{j}xF^{\prime }(x),\quad A_{j}=\tilde{A}_{j}+%
\tilde{B}_{j}D/(2\alpha ),\quad B_{j}=\tilde{B}_{j}/\alpha ,
\label{notbar}
\end{equation}%
and for a given function $g(v)$ the functions $F^{(i)}[g(v)]$ are
given by
\begin{eqnarray}
F^{(i)}[g(v)] &=&\left( \frac{1}{2}-2\xi \right) \left[
v^{2}g^{\prime 2}(v)+\left( D+\frac{4\xi }{4\xi -1}\right)
vg(v)g^{\prime }(v)+\right.
\notag \\
&&+\left. \left( \sigma ^{2}+v^{2}+\frac{2v^{2}}{D(4\xi -1)}\right) g^{2}(v)%
\right] ,\quad i=0,1,\ldots ,D-1,  \label{Finew} \\
F^{(D)}[g(v)] &=&-\frac{v^{2}}{2}g^{\prime
}{}^{2}(v)+\frac{D}{2}\left( 4\xi
-1\right) vg(v)g^{\prime }(v)+  \notag \\
&&+\frac{1}{2}\left[ \sigma ^{2}+v^{2}+2\xi D(D+1)-D^{2}/2\right]
g^{2}(v). \label{FDnew}
\end{eqnarray}%
We assume values of the curvature coupling parameter for which
$\sigma $ is
real. For imaginary $\sigma $ the ground state becomes unstable \cite{Brei82}%
. Note that VEVs (\ref{Tik1plnew2}) depend only on the ratio
$z/z_{a}$ which
is related to the proper distance from the plate by the equation $%
z/z_{a}=e^{(y-a)/\alpha }$. As we see, for the part of the
energy-momentum tensor corresponding to the coordinates in the
hyperplane parallel to the plates one has $\langle T_{\mu \nu
}\rangle ^{(a)}\propto \eta _{\mu \nu }$. Of course, we could
expect this result from the problem symmetry.

For a conformally coupled massless scalar $\sigma =1/2$, and it
can be seen that $\langle T_{i}^{k}\rangle ^{(a)}=0$ in the region
$z>z_{a}$ and
\begin{equation}
\langle T_{D}^{D}\rangle ^{(a)}=-D\langle T_{0}^{0}\rangle ^{(a)}=-\frac{%
\alpha ^{-D-1}(z/z_{a})^{D+1}}{(4\pi )^{D/2}\Gamma (D/2)}\int_{0}^{\infty }%
\frac{t^{D}\,dt}{\frac{B_{a}(t-1)+2A_{a}}{B_{a}(t+1)-2A_{a}}e^{t}+1}
\label{conf1pl}
\end{equation}%
in the region $z<z_{a}$. Note that the corresponding
energy-momentum tensor for a single Robin plate in the Minkowski
bulk vanishes \cite{Rome02} and the result for the region
$z>z_{a}$ is obtained by a simple conformal transformation from
that for the Minkowski case. In the region $z<z_{a}$ this
procedure does not work as in the AdS problem one has $0<z<z_{a}$
instead of $-\infty <z<z_{a}$ in the Minkowski problem and, hence,
the part of AdS under consideration is not a conformal image of
the corresponding manifold in the Minkowski spactime.

In the limit $z\rightarrow z_{a}$ for a fixed $\alpha $ the expressions (\ref%
{Tik1plnew2}) diverge. Near the brane, to the leading order, we have $%
\langle T_{0}^{0}\rangle ^{(a)}\propto |1-z_{a}/z|^{-D-1}$ and
$\langle T_{D}^{D}\rangle ^{(a)}\approx (1-z_{a}/z)\langle
T_{0}^{0}\rangle ^{(a)}$. In this region, the vacuum energy
densities have opposite signs for the cases of Dirichlet
($B_{a}=0$) and non-Dirichlet ($B_{a}\neq 0$) boundary conditions.
For a conformally coupled massless scalar the vacuum
energy-momentum tensor vanishes in the region $z>z_{a}$ and it is
given by the expression (\ref{conf1pl}) in the region $z<z_{a}$.
The latter is finite everywhere including the points on the plate.

For points with the proper distances from the brane much larger
compared with the AdS curvature radius one has $z\gg z_{a}$. This
limit is important from the point of view of the application to
the Randall-Sundrum braneworld. To the leading order one has
$\langle T_{0}^{0}\rangle ^{(a)}\propto (z_{a}/z)^{2\sigma }$,
$\langle T_{D}^{D}\rangle ^{(a)}\approx D\langle T_{0}^{0}\rangle
^{(a)}/(D+2\sigma )$, and the brane-induced VEVs are suppressed by
the factor $\exp [2\sigma (a-y)/\alpha ]$. The limit $z\ll z_{a}$
for a fixed $\alpha $ corresponds to points near the AdS boundary
$z=0 $, with the proper distances from the brane much larger
compared with the AdS curvature radius. In this limit to the
leading order we have $\langle T_{0}^{0}\rangle ^{(a)}\propto
(z/z_{a})^{D+2\sigma }$, $\langle T_{D}^{D}\rangle ^{(a)}\approx
-D\langle T_{0}^{0}\rangle ^{(a)}/(2\sigma )$ and the exponential
suppression is by the factor $\exp [(D+2\sigma )(y-a)/\alpha ]$.

\subsection{Two-branes geometry}

In this section we will investigate the VEV for the
energy-momentum tensor in the region between two parallel branes.
This VEV is presented in the form
\begin{eqnarray}
\langle 0|T_{i}^{k}|0\rangle &=&\langle T_{i}^{k}\rangle _{\mathrm{AdS}%
}+\langle T_{i}^{k}\rangle ^{(j)}-\frac{\alpha ^{-D-1}z^{D}\delta _{i}^{k}}{%
2^{D-1}\pi ^{D/2}\Gamma \left( D/2\right) }  \notag \\
&&\times \int_{0}^{\infty }duu^{D-1}\Omega _{j\sigma
}(uz_{a},uz_{b})F^{(i)} \left[ G_{\sigma }^{(j)}(uz_{a},uz)\right]
,  \label{Tik1int}
\end{eqnarray}%
with the functions $F^{(i)}[g(v)]$ from Eqs.
(\ref{Finew}),(\ref{FDnew}) and
with%
\begin{eqnarray}
\Omega _{a\sigma }(u,v) &=&\frac{\bar{K}_{\sigma
}^{(b)}(v)/\bar{K}_{\sigma
}^{(a)}(u)}{\bar{K}_{\sigma }^{(a)}(u)\bar{I}_{\sigma }^{(b)}(v)-\bar{K}%
_{\sigma }^{(b)}(v)\bar{I}_{\sigma }^{(a)}(u)},  \nonumber \\
\Omega _{b\sigma }(u,v) &=&\frac{\bar{I}_{\sigma
}^{(a)}(u)/\bar{I}_{\sigma
}^{(b)}(v)}{\bar{K}_{\sigma }^{(a)}(u)\bar{I}_{\sigma }^{(b)}(v)-\bar{K}%
_{\sigma }^{(b)}(v)\bar{I}_{\sigma }^{(a)}(u)},  \label{Omb} \\
G_{\sigma }^{(a)}(u,v) &=&I_{\sigma }(v)\bar{K}_{\sigma
}^{(a)}(u)-K_{\sigma }(v)\bar{I}_{\sigma }^{(a)}(u),  \nonumber
\end{eqnarray}%
where the barred notations are defined in (\ref{notbar}). In
(\ref{Tik1int}), $\langle T_{i}^{k}\rangle ^{(j)}$ is the VEV of
the energy-momentum tensor induced by a single brane located at
$y=j$. The cases $j=a$ and $j=b$ provide two equivalent
representations for the VEV.

The vacuum force acting per unit surface of the brane at $z=z_{j}$
is determined by the ${}_{D}^{D}$ -- component of the vacuum
energy-momentum tensor at this point. The corresponding effective
pressures are decomposed as
$p^{(j)}=p_{1}^{(j)}+p_{{\mathrm{(int)}}}^{(j)}$, $j=a,b$, where
the first term on the right is the pressure for a single brane at
$z=z_{j}$ when the second plate is absent. This term is divergent
due to the surface divergences and needs an additional
renormalization. The term
\begin{eqnarray}
p_{{\mathrm{(int)}}}^{(j)} &=&\frac{\alpha ^{-D-1}}{2^{D}\pi
^{D/2}\Gamma \left( D/2\right) }\int_{0}^{\infty
}dx\,x^{D-1}\Omega _{j\sigma }\left(
xz_{a}/z_{j},xz_{b}/z_{j}\right)  \notag \\
&&\times \left\{ \left( x^{2}-\sigma ^{2}+2m^{2}\alpha ^{2}\right)
B_{j}^{2}-D(4\xi -1)A_{j}B_{j}-A_{j}^{2}\right\} .  \label{pintj}
\end{eqnarray}%
is the pressure induced by the presence of the second plate, and
can be termed as an interaction force. The effective pressures
(\ref{pintj}) are negative for Dirichlet scalar and for a scalar
with $A_{a}=A_{b}=0$. In these cases the corresponding interaction
forces are attractive for all values of the branes separation.

At small distances between the branes, $(b-a)\ll \alpha $, the
leading terms
are the same as for parallel plates in the Minkowski bulk with $p_{{\mathrm{%
(int)}}}^{(j)}\propto (b-a)^{-D-1}$. In this limit the interaction
forces are repulsive in the case of Dirichlet boundary condition
on one plate and non-Dirichlet boundary condition on the another
and the forces are
attractive for all other cases. For large distances between the branes, $%
(b-a)\gg \alpha $, one has $p_{{\mathrm{(int)}}}^{(a)}\propto \exp
[(D+2\sigma )(a-b)/\alpha ]$ and
$p_{{\mathrm{(int)}}}^{(b)}\propto \exp [2\sigma (a-b)/\alpha ]$.
With dependence of the values for the coefficients in the boundary
conditions, the corresponding forces can be either repulsive or
attractive.

The results obtained in this section can be directly applied to
the Randall-Sundrum braneworld model \cite{Rand99a}. In the
$D$-dimensional generalization of this model the coordinate $y$ in
(\ref{metric}) is compactified on an orbifold, $S^{1}/Z_{2}$ of
length $L$, with $-L\leqslant y\leqslant L$. The orbifold fixed
points at $y=0$ and $y=L$ are the locations of two $(D-1)$-branes.
The metric in the Randall-Sundrum model has the form
(\ref{metric}) with $e^{-2y/\alpha }\rightarrow e^{-2|y|/\alpha
}$. The action functional for a scalar $\varphi $ has the form
\begin{equation}
S=\frac{1}{2}\int d^{D}xdy\sqrt{|g|}\left\{ g^{ik}\partial
_{i}\varphi
\partial _{k}\varphi -\left[ m^{2}+c_{1}\delta (y)+c_{2}\delta (y-L)+\xi R%
\right] \varphi ^{2}\right\} ,  \label{actionsc}
\end{equation}%
with $c_{1}$ and $c_{2}$ being the mass terms on the branes $y=0$
and $y=L$ respectively. The VEV of the energy-momentum tensor of
this scalar is obtained from the formula (\ref{Tik1int}) by an
additional factor 1/2. For
the coefficients in the boundary conditions one has (see also Ref. \refcite%
{Flac01b} for the case $c_{1}=c_{2}=0$)
\begin{equation}
\tilde{A}_{a}/\tilde{B}_{a}=-(c_{1}+4D\xi /\alpha )/2,\quad \tilde{A}_{b}/%
\tilde{B}_{b}=-(-c_{2}+4D\xi /\alpha )/2,  \label{AtildeRS}
\end{equation}%
for an untwisted scalar and $\tilde{B}_{a}=\tilde{B}_{b}=0$
(Dirichlet boundary conditions) for a twisted scalar. The
energy-momentum tensor in the
Randall-Sundrum braneworld for a bulk scalar with zero brane mass terms $%
c_{1}$ and $c_{2}$ is considered in Ref. \refcite{Knap03}. In this
paper only a general formula is given for the unrenormalized VEV
in terms of the differential operator acting on the Green
function. In our approach the part due to the AdS bulk without
boundaries is explicitly extracted and in this way the
renormalization is reduced to that in the boundary-free AdS
spacetime. In addition, we have presented the brane-induced parts
in terms of exponentially convergent integrals convenient for
numerical calculations.

We have considered the bulk energy-momentum tensor. For a scalar
field on manifolds with boundaries in addition to this part, the
energy-momentum tensor contains a contribution located on the
boundary (for the expression of the surface energy-momentum tensor
in the case of arbitrary bulk and boundary geometries see Ref.
\refcite{Saha03}). The vacuum expectation value of the surface
energy-momentum tensor for the geometry of two parallel branes in
AdS bulk is evaluated in Ref. \refcite{Saha04b}. It is shown that
for large distances between the branes the induced surface
densities give rise to an exponentially suppressed cosmological
constant on the branes. In particular, in the Randall-Sundrum
model the cosmological constant generated on the visible brane is
of the right order of magnitude with the value suggested by the
cosmological observations. The local vacuum densities in
higher-dimensional braneworld models with compact internal
subspaces have been investigated in Ref. \refcite{Saha06} (for the
total vacuum energy in these models see Ref. \refcite{Flac03}).

\section{Conclusion}

\label{sec:Conc}

In the present paper we have considered the Casimir densities for
a massive scalar field in $(D+1)$-dimensional dS spacetime having
the spatial topology $\mathrm{R}^{p}\times (\mathrm{S}^{1})^{q}$.
Both cases of the periodicity and antiperiodicity conditions along
the compactified dimensions are discussed. We have decomposed the
VEV of the energy-momentum tensor into the uncompactified dS and
topological parts. Since the toroidal compactification does not
change the local geometry, in this way the renormalization of the
energy-momentum tensor is reduced to that for uncompactifeid
$\mathrm{dS}_{D+1}$.

At early stages of the cosmological expansion the vacuum
energy-momentum tensor coincides with the corresponding quantity
for a conformally coupled massless field and the topological part
behaves like $e^{-(D+1)t/\alpha }$. In this limit the topological
part dominates in the VEV. At late stages of the cosmological
expansion and for real $\nu $ the topological part is suppressed
by the factor $e^{-(D-2\nu )t/\alpha }$ and the vacuum stresses
are isotropic. In the same limit and for pure imaginary $\nu $ the
topological terms oscillate with the amplitude going to the zero
as $e^{-Dt/\alpha }$ for $t\rightarrow +\infty $.

In the second part, we have investigated the VEV of the
energy-momentum tensor for a scalar field satisfying Robin
boundary conditions on two parallel branes in AdS spacetime. The
part of this tensor corresponding to the components on the
hyperplane parallel to the brane is proportional to the
corresponding metric tensor. The vacuum forces acting on the
branes are decomposed into self-action and interaction parts. The
interaction forces are given by formula (\ref{pintj}) with $j=a,b$
for the brane at $z=z_{a}$ and $z=z_{b}$ respectively. For
Dirichlet scalar they are always attractive. In the case of
general Robin boundary conditions the interaction forces can be
either repulsive or attractive. Moreover, there is a region in the
space of Robin parameters in which the interaction forces are
repulsive for small distances and are attractive for large
distances. This provides a possibility for the stabilization of
the interbrane distance by using the vacuum forces. For large
distances between the branes, the vacuum interaction forces per
unit surface are exponentially suppressed by the factor $\exp
[2\sigma (a-b)/\alpha ]$ for the brane at $y=a$ and by the factor
$\exp [(2\sigma +D)(a-b)/\alpha ]$ for the brane at $y=b$.

\section*{Acknowledgments}

The author acknowledges the Organizers of the 5th International
Workshop on Astronomy and Relativistic Astrophysics (IWARA2011)
and CAPES (Brazil) for a support.

\end{document}